\documentclass[journal]{IEEEtran}

\usepackage{epsfig,amsmath,amssymb,epsf,amsthm,scalefnt,multirow,dsfont,subfig}
\usepackage{xcolor}
\usepackage{float}
\usepackage{cite}

\newcommand{\argmax}{\mathop{\mathrm{argmax}}}
\newcommand{\argmin}{\mathop{\mathrm{argmin}}}

\newtheorem{lemma}{Lemma}

  \def\cC{{\mathcal{C}}} \def\cD{{\mathcal{D}}}
   
\def\cI{{\mathcal{I}}}   
 \def\cN{{\mathcal{N}}}  
  \def\cS{{\mathcal{S}}} 
\def\cU{{\mathcal{U}}}  \def\cW{{\mathcal{W}}}

\def\argmin{\mathop{\mathrm{argmin}}}
\def\argmax{\mathop{\mathrm{argmax}}}

\def\Re{\mathop{\mathrm{Re}}}
\def\Im{\mathop{\mathrm{Im}}}

\def\b0{{\pmb{0}}} 

\def\ba{{\mathbf{a}}} \def\bb{{\mathbf{b}}} \def\bc{{\mathbf{c}}} 
  \def\bg{{\mathbf{g}}} \def\bh{{\mathbf{h}}}

\def\bu{{\mathbf{u}}} \def\bv{{\mathbf{v}}}  
\def\by{{\mathbf{y}}} \def\bz{{\mathbf{z}}}

  \def\bG{{\mathbf{G}}} 
\def\bI{{\mathbf{I}}}

\newcommand{\C}{\mathbb{C}}

\begin{document}

\title{Coded Distributed Diversity: A Novel Distributed Reception Technique for Wireless Communication Systems}

\author{Junil Choi$^*$, David J. Love, and Patrick Bidigare\\
\thanks{This work has been submitted to the IEEE for possible publication. Copyright may be transferred without notice, after which this version may no longer be accessible.}
\thanks{Junil Choi and David Love are with the School of Electrical and Computer Engineering, Purdue University, West Lafayette, IN (e-mail: junil.choi@gmail.com, djlove@ecn.purdue.edu).}
\thanks{Patrick Bidigare is with Raytheon BBN Technologies, Arlington, VA (e-mail: bidigare@ieee.org).}
\thanks{This paper was presented in part at the Asilomar Conference on Signals, Systems, and Computers, 2013 \cite{love_asilomar}.}
}


\maketitle

\begin{abstract}
In this paper, we consider a distributed reception scenario where a transmitter broadcasts a signal to multiple geographically separated receive nodes over fading channels, and each node forwards a few bits representing a processed version of the received signal to a fusion center.  The fusion center then tries to decode the transmitted signal based on the forwarded information from the receive nodes and possible channel state information.  We show that there is a strong connection between the problem of minimizing a symbol error probability at the fusion center in distributed reception and channel coding in coding theory.  This connection allows us to design a unified framework for coded distributed diversity reception.  We focus linear block codes such as simplex codes or first-order Reed-Muller codes that achieve the Griesmer bound with equality to maximize the diversity gain.  Due to its simple structure, no complex offline optimization process is needed to design the coding structure at the receive nodes for the proposed coded diversity technique.  The proposed technique can support a wide array of distributed reception scenarios, i.e., arbitrary $M$-ary symbol transmission at the transmitter and received signal processing with multiple bits at the receive nodes.  Numerical studies show that the proposed coded diversity technique can achieve practical symbol error rates even with moderate signal-to-noise ratio and numbers of the receive nodes.
\end{abstract}


\section{Introduction}\label{sec1}
Distributed transmission and/or reception have become popular in many wireless signal processing scenarios including cellular systems, target detection in radar systems, wireless sensor networks, and military communications.  Coordinated multipoint (CoMP) in the 3GPP standard \cite{comp2,comp1,comp3} enables multiple base stations to cooperate with each other to support cell edge users by joint transmission (JT) \cite{jp1,jp2} or coordinated scheduling/coordinated beamforming (CS/CB) \cite{cscb1,cscb2}.  Distributed antenna systems (DAS) are also adopted to boost performance in cellular systems \cite{das1,das2,das3}.  The geographically separated radio entities in radar systems can obtain different information of a target (or multiple targets) and make better decisions, e.g., location or speed of the target \cite{radar1,radar2,radar3}.  In wireless sensor networks, transmission/reception techniques are even more crucial because sensors are usually very cheap and only can perform simple operations \cite{wsn1,wsn2,wsn3,wsn4}.  Military communications where a squad of radio units serves as a distributed array in battlefields can be considered as a form of distributed multiple antenna systems \cite{military1,military2}.

We focus on distributed reception \cite{dist_detect1,dist_detect2,dist_detect3} to provide diversity advantage in fading channels in this paper.  We assume that there is a transmitter that wants to send a signal to a fusion center by the help of multiple geographically separated receive nodes.  Each node receives the broadcasted signal from the transmitter through fading channels and forwards the \textit{processed} received signal to the fusion center.  The fusion center then tries to decode the transmitted signal using the forwarded information from the receive nodes and, if available, channel state information (CSI).

This scenario has been studied in \cite{dist_detect2} and \cite{brown} for cases when the number of processing bits at each receive node is greater than or equal to the number of bits representing data symbol constellation. Our focus is on more practical case when each receive node \textit{quantizes} the received signal before forwarding it to the fusion center.  This scenario is of particular interest when the number of receive nodes is \textit{large} because the data rate of the link from each receive node to the fusion center might be constrained.

We show that there is a strong connection between the problem of minimizing a symbol error probability at the fusion center in distributed reception and channel coding in coding theory.  In coding theory, we achieve time diversity in fading channels by transmitting channel coded data bits using multiple channel instances \cite{fading}.  Similarly, we can obtain spatial diversity by exploiting multiple receive nodes that experience weakly correlated or independent channels in distributed reception.  This connection allows us to utilize well-established channel coding techniques to develop good distributed reception strategies and achieve the maximum diversity gain.  The achieved diversity gain by distributed reception would give range and/or data rate advantages.

The connection between the distributed reception problem and channel coding has been first explored in \cite{code_dist_detec1,code_dist_detec2,code_dist_detec3} for the distributed fault-tolerant classification problem in wireless sensor networks.  A codeword set matrix is generated by two algorithms, i.e., cyclic column replacement and simulated annealing, for single bit and multiple bits receive node processing in \cite{code_dist_detec1} and \cite{code_dist_detec2}, respectively.  Each codeword (or a symbol in a signal constellation set) forms a row in the codeword set matrix and each column of the matrix represents the decision rule employed at each receive node.  The proposed approaches in \cite{code_dist_detec1,code_dist_detec2}; however, are heuristic and do not guarantee optimality in any sense.  Moreover, those approaches need complex offline optimization to generate code matrices for every different number of the receive nodes.

In this paper, we consider three general scenarios: 1) fading channels between the transmitter and the receive nodes, 2) arbitrary $M$-ary data symbol transmission from the transmitter, and 3) multiple bits processing at the receive nodes.  To support these scenarios effectively, we propose a unified framework of processing at the receive nodes and decoding at the fusion center.  We dub the unified framework a \textit{coded diversity technique}.  The coded diversity technique fully exploits the connection of the distributed reception problem with coding theory, and we are able to exploit efficient linear block codes such as simplex codes or first-order Reed-Muller codes that achieve the Griesmer bound with equality.  We also develop novel shortened concatenated repetition-simplex (SCRS) codes for an arbitrary number of the receive nodes and show that the SCRS codes are optimal with respect to the Griesmer bound in many practical scenarios.  The SCRS codes are very easy to generate such that we do not need to perform any kind of complex optimization process to generate a SCRS code for an arbitrary number of receive nodes.  We show the performance of the proposed coded diversity technique by analytical derivations and numerical studies as well.

This paper is organized as follows.  We show a motivating example of this work and explain a general system model in Section \ref{sec2}.  In Section \ref{sec4}, we explain the general framework and the diversity order analyses of the proposed coded diversity technique.  Practical code designs and their performance implications for the coded diversity technique are explained in Section \ref{sec5}.  Numerical studies are presented in Section VI, and conclusions follow in Section VII.
\vspace{0.2cm}

\section{Motivating Example and System Model}\label{sec2}
We show a motivating example of this work and explain a general system model.

\subsection{Motivating Example}
Consider single-input multiple-output (SIMO) system with three geographically separated receive nodes.  The $i$th receive node operates with an input-output equation
\begin{equation}\notag
y_i = h_i s + n_i,\quad i=1,2,3
\end{equation}
with $h_i\in \mathbb{C}$ is a channel from the transmitter to the node $i$, $s\in\mathcal{S}\subset\C$ is the transmitted signal selected from $\cS$ with a uniform distribution, and $n_i$ is complex additive white Gaussian noise (AWGN) distributed as $\cC\cN(0,1).$  We assume that the noise is spatially independent, i.e., each receive node experiences independent noise.  We further assume that the channel collected across the distributed array $\bh = [h_1~h_2~h_3]$  is a spatially uncorrelated channel with $\cC\cN(0,1)$ entries and the $i$th receive node has perfect knowledge of $h_i$ and no knowledge of the other users' channels.

If the fusion center knows $\bh = [h_1~h_2~h_3]$ and $\by = [y_1 ~y_2~y_3]$ perfectly, then as in a standard, centralized combining system, the fusion center can produce
\begin{equation}\label{cas}
\widetilde{y} = \frac{\mathbf{y}\mathbf{z}^*}{\bh\bz^*}
\end{equation}
where $\bz=\bh/\|\bh\|$ is the optimal linear combiner.  The processed output $\widetilde{y}$ is used to detected the transmitted symbol $s$.  However, the main focus of this paper is the case when each receive node only can send the processed version of $y_i$,  which we denote $u_i$ throughout the paper, using a small number of bits per channel use to the fusion center, and the fusion center tries to decode the transmitted symbol based on $u_i$'s along with possibly the knowledge of $\bh$.  We assume that each node can forward $u_i$ without any error to the fusion center.\footnote{This assumption is reasonable for many scenarios, e.g., 1) the receive nodes are connected with the fusion center through wired lines as in CoMP, DAS, or radar systems, 2) the receive nodes and the fusion center are closely located with each other in wireless sensor networks.}  Many receive architectures in both commercial and military systems fall into this distributed reception scenario.

In this example, we focus on the case when each node can pass only \textit{one bit} for $u_i$ per channel use to the fusion center; however, the transmitted symbol $s$ is uniformly selected from a quadrature phase shift keying (QPSK)
\begin{equation}\notag
\mathcal{S} = \left\{\sqrt{\frac{\rho}{2}}(1+j),\sqrt{\frac{\rho}{2}}(1-j),\sqrt{\frac{\rho}{2}}(-1+j),\sqrt{\frac{\rho}{2}}(-1-j)\right\}
\end{equation}
where $\rho$ denotes the transmit signal-to-noise ratio (SNR).  Thus, the fusion center needs to detect the transmitted symbol using 3 bits (1 bit per receive node) per channel use.

With a naive approach, this problem can be mapped into a binary hypothesis testing problem at each node.  For example, nodes 1 and 3 detect the real component such that
\begin{equation*}
  u_i=
  \begin{cases}
    1 & \text{if}~~ \Re(y_i)\geq 0 \\
    0 & \text{if}~~ \Re(y_i)< 0
  \end{cases}, \quad i=1,3
\end{equation*}
while node 2 detects the imaginary component as
\begin{equation*}
  u_2=
  \begin{cases}
    1 & \text{if}~~ \Im(y_2)\geq 0 \\
    0 & \text{if}~~ \Im(y_2)< 0
  \end{cases},
\end{equation*}
and send their decisions $u_i$ to the fusion center.  With an assumption that each node $i$ has perfect knowledge of its channel $h_i$, a probability of incorrectly detecting the desired component at node $i$ is given by
\begin{equation}\label{pi_awgn}
  P_e(h_i,\rho) =Q\left(\sqrt{|h_i|^2\rho}\right)
\end{equation}
for $i=1,2,3$.  With full CSI knowledge at the fusion center, maximum likelihood (ML) detection will give a probability of symbol error as
\begin{equation*}
P_{e,unc}(\bh,\rho) = 1-\left(1-\max_{i\in\{1,3\}}P_e(h_i,\rho)\right)\left(1-P_e(h_2,\rho)\right).
\end{equation*}
Let $P_{e,unc}(\rho) = E\left[P_{e,unc}(\bh,\rho) \right]$ where expectation is taken over $\bh$.  Then, the diversity order is given as
\begin{equation}\label{def_diversity}
-\lim_{\rho\rightarrow\infty}\frac{\log(P_{e,unc}(\rho))}{\log\rho} = 1,
\end{equation}
meaning that the receiver can only get the diversity order 1.  This is a discouraging result because the distributed reception with three nodes has not provided any increase in diversity.

Note that a better solution exists.  As in the previous approach, node 1 and 2 detect the real and imaginary component, respectively.  However, node 3 now detects the product of the real and imaginary components such that
\begin{equation*}
  u_3=
  \begin{cases}
    1 & \text{if}~~ \Re(y_3)\Im(y_3) \geq 0 \\
    0 & \text{if}~~ \Re(y_3)\Im(y_3) < 0
  \end{cases}.
\end{equation*}
Nodes 1 and 2 have a probability of incorrect detection as in \eqref{pi_awgn} while node 3 has a probability of detecting incorrectly given by
\begin{equation*}
  P_e(h_3,\rho)=  2Q\left(\sqrt{|h_3|^2\rho}\right)\left(1-Q\left(\sqrt{|h_3|^2\rho}\right)\right).
\end{equation*}
If we let $P_{e,(i)}(h_{(i)},\rho)$ be the $i$-th largest probability of error among the nodes such that
\begin{equation*}
P_{e,(1)}(h_{(1)},\rho) \geq P_{e,(2)}(h_{(2)},\rho) \geq P_{e,(3)}(h_{(3)},\rho),
\end{equation*}
a probability of error at the fusion center is given by
\begin{equation}\notag
P_{e,code}(\bh,\rho) = 1-\left(1-P_{e,(2)}(h_{(2)},\rho)\right)\left(1-P_{e,(3)}(h_{(3)},\rho)\right)
\end{equation}
with ML detection.  Then, the diversity order becomes
\begin{equation}\notag
-\lim_{\rho\rightarrow\infty}\frac{\log(P_{e,code}(\rho))}{\log\rho} = 2
\end{equation}
where $P_{e,code}(\rho) = E\left[P_{e,code}(\bh,\rho) \right]$ with the expectation taken over $\bh$.

Thus, without increasing the number of bits sent from any of the nodes to the fusion center or changing the channel model, we have increased the diversity order from 1 to 2 by using a smart detection scheme at each node.  More generally, this paper aims to address the following question:

\noindent ``\textit{How should each receive node quantize $y_i$ into a small number of bits to be sent to the fusion center when detecting $M$-ary modulation in distributed reception?}''

\noindent As we show later, this problem has intriguing ties to coding theory.

\subsection{System Model}
\begin{figure}[t]
\centering
 \includegraphics[width=0.6\columnwidth]{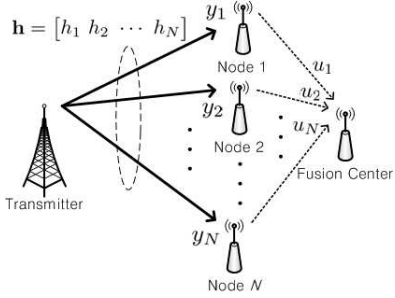}\\
  \caption{A conceptual figure of distributed reception.}\label{dist_recep}
\end{figure}
We consider a network consists of the transmitter, the fusion center, and $N$ geographically separated receive nodes.  The conceptual figure of our system model is shown in Fig. \ref{dist_recep}.  The received signal at the $i$-th node, $y_i$, is written as
\begin{equation}\notag
y_i = h_i s + n_i,\quad i=1,\cdots,N
\end{equation}
where $s\in \mathcal{S}$ is the transmitted symbol from an $M$-ary constellation
\begin{equation}\notag
\mathcal{S} = \{s_1,s_2,\ldots,s_M\}\subset\C.
\end{equation}
We assume $h_i$ and $n_i$ have the same distributions as in the motivating example.  We further assume that $s$ is selected from $\mathcal{S}$ with an equal probability and satisfies $E[s]=0$ and $E[|s|^2]=\rho$.  We define the symbol error probability at the $i$-th receive node as
\begin{equation}\label{sep_node}
  P_e(h_i,\rho) \triangleq \frac{1}{M} \sum_{s^* \in \mathcal{S}}Pr\left(s^* \neq \widehat{s} \mid s=s^*,h_i, \rho \right).
\end{equation}

Note that the majority of the distributed reception work has been dedicated for binary modulation schemes, i.e., binary hypothesis testing in AWGN channels without fading.  In this case, $s\in\mathcal{S}=\{s_1, s_2\}$, and each node can make a hard decision on the transmitted symbol.  We consider generalized distributed reception in this paper such that the transmitter can send the symbol from an arbitrary $M$-ary constellation, and each node also can send multiple bits to the fusion center.

\section{Coded Diversity and Diversity Order}\label{sec4}
We first explain the general concept of coded distributed diversity and then discuss the symbol detection schemes using the quantized node information.  We finish this section with diversity order analyses with respect to the decoding schemes.

\subsection{General Concept of Coded Diversity Technique}
Note that the $M$-ary constellation $\cS$ can be represented with a $\log_2(M) $-bit message that we denote as $\mathbf{b} = [b_1~b_2\cdots b_{\log_2(M)}]$.  Each node quantizes its received signal $y_i$ into a $B$-bit vector\footnote{We let $GF(q)^m$ denote the $m$-dimensional vector of elements in $GF(q).$  This is different from $GF(q^m)$ which denotes the finite field of order $q^m$.} $u_i \in GF(2^B)$.  We assume
\begin{equation*}
  B \leq \log_2(M)
\end{equation*}
to limit the overhead needed for the distributed decisions.  This gives rise to the concept of a \textit{compression ratio} that is defined as
\begin{equation}\notag
K  = \frac{\log_2(M)}{B}
\end{equation}
which satisfies $K\geq 1$.  We assume $K$ is an integer value throughout the paper.  We let $\ba  =[a_1 ~a_2~\cdots a_K]$ be the vectorized version of $\mathbf{b}$ with entries in $GF(2^B)$.  There are multiple ways of converting $\bb$ into $\ba$ using different primitive polynomials of $GF(2^B)$; however, using a specific primitive polynomial does not affect average performance.

Commonly detectors for $M$-ary constellations are designed using  non-overlapping decisions regions.  Denote the decision regions by $\{\cW_1,\ldots,\cW_M\}$  such that
\begin{equation*}
  \cW_{1}\bigcup\cdots\bigcup\cW_{M}=\mathbb{C}.
\end{equation*}
Using the decision regions, the detection problem can be formulated as
$
\widehat{s} = s_{m_0}
$ 
with
\begin{equation}\notag
m_0 =\argmax_{1\leq m\leq M} \mathds{1}\left(\widetilde{y}\in\cW_{m}\right)
\end{equation}
with $\mathds{1}(\cdot) $ denoting the indicator function which returns $1$ if the argument is true.  This can be also rewritten as
\begin{equation}\notag
\widehat{s} = \argmin_{s'\in\cS} \|\widetilde{y} - s'\|^2.
\end{equation}

In our problem; however, we assume that the number of decision regions at each receive node is smaller than $M$, i.e., the compression ratio is constrained as $K\geq 1$.  Let the non-overlapping decision regions at node $i$ be $\left\{\cD_{i,1},\ldots, \cD_{i,2^B}\right\}$ such that the union of the regions spans the complex plane.  The distributed reception problems can be succinctly stated as determining the sets of decision regions $\left\{\cD_{i,1},\ldots, \cD_{i,2^B}\right\}$ for $i=1,\ldots, N$ to minimize the probability of symbol detection error at the fusion center.  As shown in the motivating example of Section \ref{sec2}, this problem is nontrivial.  We show how well-developed coding techniques can be used to design the sets of decision regions.

Because of the constraint on the compression ratio $K\geq 1$, we assume the decision regions $\left\{\cD_{i,1},\ldots, \cD_{i,2^B}\right\}$ of node $i$ are constructed by certain linear combinations of the constellation decision regions $\left\{\cW_{1},\ldots,\cW_M\right\}$.  To do this, we formulate the problem using finite field notations.

To simplify the notation, let $\bb$ denote the bit representation of the the transmitted symbol $s$.  Suppose that the node $i$ first performs a hard decision on the received signal $y_i$ to generate a $\log_2(M)$-bit vector $\widehat{\bb}_i$.  If the node $i$ detects $s$ correctly, then we have $\widehat{\bb}_i=\bb$.  Note that $\widehat{\bb}_i$ can be represented with a $K$-entry vector $\widehat{\ba}_i$ with entries in $GF(2^B)$.  The node $i$ then generates $u_i = f_i\left(\widehat{\ba}_i\right)$ using a function
\begin{equation*}
    f_i: ~GF(2^B)^{K}\rightarrow GF(2^B)
\end{equation*}
and sends $u_i$ to the fusion center.

If each node received $\ba$ (or equivalently $\bb$) without any error, this problem can be formulated as a coding problem.  The $K$-dimensional \textit{message} $\ba$ is transformed to an $N$-dimensional vector codeword $\bu = [u_1~\cdots~u_N]$ with entries in $GF(2^B).$    In the distributed reception case, each node is coding on noisy data, and the vector $\bu$ is corrupted with \textit{noise} corresponding to reception error at each node.  Despite this, the goal in distributed reception is very similar to code design in coding theory.  We must find a coding technique that minimizes decoding error of the transmitted symbol at the fusion center.

Similar to the coding problem, we focus on the creation of an $M$ vector codeword set $\{\bu[1],\ldots, \bu[M]\}$ where each codeword $\bu[k]$ corresponds to a constellation point $s_k\in \cS$.  Further, we focus on linear block codes to enable efficient encoding.  This means that the function $f_i$ is explicitly given as
\begin{equation}\label{node_process}
u_i=f_i\left(\widehat{\ba}_i\right) = \widehat{\ba}_i \mathbf{g}_i^T
\end{equation}
where $\bg_i\in GF(2^B)^{K}.$  We can collect everything together in vector form such that
\begin{equation}\notag
\bu = \bc(s) + \bv
\end{equation}
where $\bc(s)\in GF(2^B)^N$ denotes the distributed detection bits if all nodes make the correct bit decisions when $s$ is transmitted and $\bv\in GF(2^B)^N$ represents \textit{noise} caused by reception error at each node.  Due to the linear structure, a generator matrix $\bG\in GF(2^B)^{K\times N}$ can be given as
\begin{equation}\notag
\bG = \left[\begin{array}{ccc}
\bg_1  \\
\vdots  \\
\bg_N
\end{array}
\right]^T
\end{equation}
and
\begin{equation}\notag
\bc(s) = \ba\bG.
\end{equation}
This generates a code for the constellation points $\cS$ as\footnote{This code generation method is completely different from the one proposed in \cite{code_dist_detec1}.  We briefly explain the scheme from \cite{code_dist_detec1} and compare it with the proposed coded diversity technique in Section \ref{sec6}.}
\begin{equation}\label{proposed_code_matrix}
    \cC = \{\bc(s):s\in\cS\}.
\end{equation}
We can also define the minimum Hamming distance of the code
\begin{equation}\notag
d_{\min}(\cC) = \min_{s,s'\in\cS:s\neq s'}d_{H}(\bc(s),\bc(s'))
\end{equation}
where $d_{H}(\cdot,\cdot)$ denotes the Hamming distance metric.

To explain the procedure of the coded diversity technique in words, each receive node $i$ processes $\widehat{\ba}_i$ (which is nothing but a hard-detected version of $y_i$) with the $i$-th column of $\bG$ as in \eqref{node_process}.  With $\bu = [u_1~\cdots~u_N]$ from all $N$ receive nodes, the fusion center detects the transmitted symbol by using decoding schemes explained next.

\subsection{Decoding Schemes at Fusion Center}\label{decoding_schemes}
For practical reasons, we assume that the $i$-th node has knowledge \textit{only} of $\cS$, $y_i$, and $h_i$.  Each node passes $B$-bit vector $u_i$ to the fusion center, and the fusion center tries to detect the transmitted symbol $s\in \cS$ using $\bu = [u_1~\cdots~u_N]$.  We consider the cases when the fusion center has knowledge of $\bh$ and lacks knowledge of $\bh$.  Note that the fusion center does not have full access to $\by$ in our scenario, which prevents the use of the linear combiner $\bz = \mathbf{h}/\|\bh\|$ to estimate $\widetilde{y}$ as in \eqref{cas} even with full knowledge of $\bh$.  We discuss three different decoding schemes for distributed reception over fading channels:

\vspace{0.3cm}
\noindent \textbf{1) ML decoding with full CSI:} If the fusion center has full access to $\bh$, it computes
\begin{align*}
\widehat{s} &= \argmax_{s\in\cS}  Pr\left(u_1,\ldots, u_N\mid s,\bh\right)\\
  & =  \argmax_{s\in\cS} \prod_{i=1}^N Pr\left(u_i\mid s, h_i\right)
\end{align*}
where $Pr(\cdot)$ denotes the probability that is computed as
\begin{equation}\notag
Pr\left(u_i\mid s, h_i\right)  = \frac{1}{\pi} \int_{\cD_{i,u_i}} e^{
-\left|y_i - h_i s \right|^2
}dy_i
\end{equation}
with $\cD_{i,u_i}$ denoting the decision region of node $i$ corresponding to the $B$-bit pattern $u_i$.\footnote{In practice, we can generate empirical probabilities of $Pr\left(u_i\mid s, h_i\right)$ in advance to perform ML decoding with full CSI.}

\vspace{0.3cm}
\noindent \textbf{2) Selected subset ML decoding with full CSI:} If the number of receive nodes $N$ is very large, the complexity of ML decoding can be excessive.  With our coded diversity framework; however, we can reduce decoding complexity significantly while obtaining comparable performance with ML decoding.

First, we assume that every $L$-th receive node shares the same processing rule, i.e.,
\begin{equation*}
\bg_{i}=\bg_{i+L}=\cdots=\bg_{i+\left\lfloor\frac{N-i}{L}\right\rfloor L},\quad i\in\{1,\ldots,L\}
\end{equation*}
and $\bg_i\neq \bg_k$ if $i\neq k$ for $i,k\in\{1,\ldots,L\}$.  Let the set of nodes that share the same processing rule with node $i$ as $\mathcal{I}_i=\{i,i+L,\ldots,i+\left\lfloor\frac{N-i}{L}\right\rfloor  L\}$.  Because the fusion center has full access to $\bh$, it can select the node $i^{\dag}$ among $\cI_i$ as
\begin{equation*}
  i^{\dag} = \argmax_{k\in \mathcal{I}_i}|h_k|^2,\quad i=1,\ldots,L,
\end{equation*}
perform ML decoding using bits from the $L$ selected receive nodes.  This selected subset ML decoding is appropriate for one of our code design explained in Section \ref{cb_selection}.
\begin{figure*}[t]
\begin{equation}\label{pr_bound1}
Pr\left(\left.\prod_{i:c_i(s)\neq c_i(s')}\left( Pr\left(u_i\mid {c}_i(s'), h_i\right)-Pr\left(u_i\mid {c}_i(s), h_i\right) \right) > 0  \right| \bh, s, s'\right) \leq \prod_{i:c_i(s)\neq c_i(s')} P_e(h_i,\rho).
\end{equation}
\hrule
\end{figure*}

\vspace{0.3cm}
\noindent \textbf{3) Minimum Hamming distance decoding without CSI:} If the fusion center does not have any knowledge of $\bh$, it needs to rely on the simple Hamming distance decoding.  The fusion center then tries to detect the transmitted symbol as
\begin{equation*}
\widehat{s} = \argmin_{s\in\cS}  d_{H}(\bc(s),\bu)
\end{equation*}
where $\bc(s)$ is the vector that would be sent from the all $N$ receive nodes if $s$ were perfectly decoded at each node and $\bu = [u_1~\cdots~u_N]$.

\subsection{Diversity Analysis}
We present the diversity analyses of the proposed coded diversity technique with three different decoding schemes explained in the previous section in the following.

\begin{lemma}\label{lemma_soft}
A coded distributed diversity system using a codeword set $\cC = \{\mathbf{c}(s_1),\ldots,\mathbf{c}(s_M)\}$ achieves a diversity order of $d_{\min}(\cC)$ using ML decoding.
\end{lemma}
\begin{IEEEproof}
The diversity order of the probability of error $P_{e,code,ML}(\rho)$ can be obtained  by analyzing the worst case pairwise error probability
$\max\limits_{s\neq s'}Pr(s\rightarrow s').$

Instead of working with the optimal decoder directly, we will bound performance on a suboptimal decoder.
The considered suboptimal detector chooses\begin{equation}\notag
\widehat{s} =  \argmax_{s\in\cS} \prod_{i=1}^N Pr\left(u_i\mid c_i(s), h_i\right)
\end{equation}
where $c_i(s)$ is the $i$-th entry of $\bc(s).$  In the event of a tie, it is broken arbitrarily.  Using the coding framework and pairwise error probability,
\begin{equation}\notag
\prod_{i=1}^N Pr\left(u_i\mid {c}_i(s'), h_i\right)-\prod_{i=1}^N Pr\left(u_i\mid {c}_i(s), h_i\right)  > 0
\end{equation}
or
\begin{equation}\notag
\prod_{i:c_i(s)\neq c_i(s')}\left( Pr\left(u_i\mid {c}_i(s'), h_i\right)- Pr\left(u_i\mid {c}_i(s), h_i\right) \right) > 0.
\end{equation}
Then, we can have a bound as in \eqref{pr_bound1} where $P_e(h_i,\rho)$ is a probability of symbol error at node $i$ defined in \eqref{sep_node}. Taking the expectation over $h_i$, the pairwise error probability is bounded as
\begin{equation}\notag
Pr(s\rightarrow s') \leq E\left[(P_e(h,\rho))^{d_{H}(\bc(s),\bc(s'))}\right]
\end{equation}
where the expectation is over $h$.\footnote{Because we assume every node experiences i.i.d. Rayleigh fading channel, we drop the receive node index $i$.}  Therefore,
\begin{equation}\notag
\max_{s\neq s'}Pr(s\rightarrow s') \leq E\left[(P_e(h,\rho))^{d_{\min}(\cC)}\right]
\end{equation}
which gives the lower bound of the diversity order of $d_{\min}(\cC)$.

To obtain a lower bound on $Pr(s\rightarrow s')$ (or the upper bound of the diversity order), we can consider a diversity combiner that observe all
$\{y_i\}_{i:c_i(s)\neq c_i(s')}$ and compute a maximum ratio combining (MRC) combiner as
\begin{equation}\notag
\tilde{y} = \frac{\sum_{i:c_i(s)\neq c_i(s')} h_i^* y_i }{\sum_{i:c_i(s)\neq c_i(s')}|h_i|^2}.
\end{equation}
Taking the maximum over any pair $s\neq s'$ yields and average probability of error with diversity order $d_{\min}(\cC).$
\end{IEEEproof}

\begin{lemma}\label{lemma_approx}
If $L$, the number of distinctive processing rules, divides the total number of receive nodes $N$, a coded distributed diversity system using a codeword set $\cC = \{\mathbf{c}(s_1),\ldots,\mathbf{c}(s_M)\}$ achieves a diversity order of $d_{\min}(\cC)$ using selected subset ML decoding.
\end{lemma}
\begin{IEEEproof}
The diversity order of selected subset ML is upper bounded by that of ML decoding, i.e., $d_{\min}(\cC)$.  To obtain the lower bound, we again rely on the pairwise error probability.  First, we let $\bG_{[1:L]}$ be the generating matrix consists of the first $L$ columns of $\bG$ and $\cC_L = \{\mathbf{c}_L(s_1),\ldots,\mathbf{c}_L(s_M)\}$ be the resulting code from $\bG_{[1:L]}$.  Denote $d_{\min}(\cC_L)$ the minimum Hamming distance of $\cC_L$.  The symbol error probability at a group $\cI_i$ that shares the same processing rule with the $i$-th node can be given as
\begin{equation*}
  P_{e,\cI_i}(\bh_{\cI_i},\rho) = \prod_{k\in \cI_i}P_e(h_k,\rho)
\end{equation*}
where $\bh_{\cI_i}=\left[h_i,\ldots,h_{i+\left\lfloor\frac{N-i}{L}\right\rfloor L}\right]$.  Similar to the proof of Lemma \ref{lemma_soft}, the pairwise error probability is bounded as
\begin{align*}
Pr(s\rightarrow s') &\leq E\left[(P_{e,\cI}(\bh_{\cI},\rho))^{d_{H}(\bc_L(s),\bc_L(s'))}\right]\\
&=E\left[(P_{e}(h,\rho))^{\frac{N}{L}d_{H}(\bc_L(s),\bc_L(s'))}\right]
\end{align*}
where the expectation is over $h$ and the equality comes from the fact that there are $N/L$ receive nodes in $\cI$. Thus, the worst case pairwise error probability now becomes
\begin{equation}\notag
\max_{s\neq s'}Pr(s\rightarrow s') \leq E\left[(P_e(h,\rho))^{\frac{N}{L}d_{\min}(\cC_L)}\right].
\end{equation}
When $L$ divides $N$, it is obvious that $d_{\min}(\cC)=\frac{N}{L}d_{\min}(\cC_L)$ which finishes the proof.
\end{IEEEproof}

\noindent \textbf{Remark:} In general $N$ case, the diversity order of selected subset ML would lie between $\lfloor\frac{N}{L}\rfloor d_{\min}(\cC_L)$ and $\frac{N}{L}d_{\min}(\cC_L)$.

\begin{lemma}\label{lemma_hard}
A coded distributed diversity system using a codeword set $\cC = \{\mathbf{c}(s_1),\ldots,\mathbf{c}(s_M)\}$ achieves a diversity order of $\left\lceil d_{\min}(\cC)/2\right\rceil$ using minimum Hamming distance decoding.
\end{lemma}
\begin{IEEEproof}
First, we let $p = E\left[P_e(h,\rho)\right]$ to simplify notations.  Note that $0\leq p \leq 1$.  Now, consider again the pairwise error probability.
If the number of nodes with incorrect receptions is  $\lceil d_{\min}(\cC)/2\rceil$  or more, the error pattern will fall outside of the Hamming sphere of radius $\lfloor (d_{\min}(\cC)-1)/2\rfloor$ centered at the correct codeword. Thus, we have
\begin{align*}
P_{e,code,H}(\rho)&\leq
\sum_{i=\lceil d_{\min}(\cC) /2 \rceil}^{N}  {N\choose i} p^i (1-p)^{N-i}\\
&\leq N!\,\sum_{i=\lceil d_{\min}(\cC) /2 \rceil}^{N} p^i (1-p)^{N-i} \\
&\leq N!\,\sum_{i=\lceil d_{\min}(\cC) /2 \rceil}^{N} p^i \\
&\leq (N+1)!\, p^{\lceil d_{\min}(\cC) /2 \rceil}
\end{align*}
and
\begin{equation}\notag
P_{e,code,H}(\rho)
 \geq  p^{\lceil d_{\min}(\cC) /2 \rceil} (1-p)^{N-\lceil d_{\min}(\cC) /2 \rceil}.
\end{equation}
Using the fact that $p\rightarrow 0$ as $\rho \rightarrow \infty$, the diversity order is bounded as
\begin{align*}
&-\lim_{\rho\rightarrow\infty}\frac{\log(P_{e,code,H}(\rho))}{\log\rho} \\
&\qquad \geq -\lim_{\rho\rightarrow\infty}\frac{\log\left((N+1)!\, p^{\lceil d_{\min}(\cC) /2 \rceil}\right)}{\log\rho}\\
&\qquad \rightarrow \lceil d_{\min}(\cC) /2 \rceil
\end{align*}
and
\begin{align*}
&-\lim_{\rho\rightarrow\infty}\frac{\log(P_{e,code,H}(\rho))}{\log\rho} \\
&\qquad \leq -\lim_{\rho\rightarrow\infty}\frac{\log\left(p^{\lceil d_{\min}(\cC) /2 \rceil} (1-p)^{N-\lceil d_{\min}(\cC) /2 \rceil}\right)}{\log\rho} \\
&\qquad \rightarrow \lceil d_{\min}(\cC) /2 \rceil
\end{align*}
which finishes the proof.
\end{IEEEproof}

\section{Code Design and Performance Implications}\label{sec5}
Lemmas \ref{lemma_soft}, \ref{lemma_approx}, and \ref{lemma_hard} all show that the coding structure across the receive nodes dictates system performance, and it is better to have as large minimum Hamming distance as possible for a code $\cC$ (or $\cC_L$ for selected subset ML).  The coding structure is heavily dependent on the number of nodes $N$ and the compression ratio $K$.  In this section we aim to clarify this relationship and look at some simple codes that can be employed.

\subsection{Code Bounds}
The most common approaches to understand codes in coding theory are metric ball bounds, particularly  the sphere packing bound and Gilbert-Varshamov bound.  Recall that the volume of a metric ball of radius $t$ in $GF(2^B)^N$ has a volume given by
\begin{equation}\notag
V(N,t) = \sum_{ i=0}^t {N\choose i} (2^B-1)^i.
\end{equation}
The sphere packing bound requires that the $M$ balls of any code of minimum distance $d_{\min}(\cC)$   must satisfy
\begin{equation}\notag
M V\left(N,\left\lfloor \frac{d_{\min}(\mathcal{C})-1}{2}\right\rfloor\right) \leq 2^{N}.
\end{equation}
The Gilbert-Varshamov bound tells us that a linear code of minimum distance  $d_{\min}(\cC)$  exists for our $N$ node $M$-ary code if
\begin{equation}\notag
M V\left(N-1,d_{\min}(\mathcal{C})-2\right) \leq 2^{N}.
\end{equation}
These metric ball bounds are most useful in understanding code properties when $K$ grows with $N,$ particularly when $K/N$ converges to a fixed value as $N\rightarrow \infty$.  However, we are more concerned with the case where $K$ is fixed and does not scale with $N$.  Moreover, we are interested in the case when $K$ is relatively small and $N$ is not extremely large.

The most applicable bound to this situation is the Griesmer bound \cite{griesmer_bound,roth_coding}.  The Griesmer bound shows that the smallest $N$ of a code $\cC$ that can achieve a minimum Hamming distance of $d_{\min}(\cC)$ must satisfy
\begin{equation}\notag
N \geq \sum_{i=0}^{K-1} \left\lceil \frac{d_{\min}(\cC)}{2^{iB}}\right\rceil
\end{equation}
Removing the ceiling function to generate a further lower bound gives us the following bound.
\begin{lemma}
The minimum distance of any $2^B$-ary of length $N$ code $\cC$ must satisfy
\begin{equation*}
\frac{N 2^{(K-1)B}}{1 + 2^{B} + \cdots + 2^{(K-1)B}} \geq   d_{\min}(\cC).
\end{equation*}
\end{lemma}
\begin{IEEEproof}
The Griesmer bound can be lower bounded as
\begin{equation}\notag
N \geq  \left(\sum_{i=0}^{K-1}  \frac{1}{2^{iB}}\right) d_{\min}(\cC).
\end{equation}
Reformulating this by scaling both sides,
\begin{equation}\notag
\frac{N 2^{(K-1)B}}{1 + 2^{B} + \cdots + 2^{(K-1)B}} \geq   d_{\min}(\cC).
\end{equation}
\end{IEEEproof}

\subsection{Code Selection}\label{cb_selection}
The Griesmer bound gives us insight into code choice for many different scenarios (e.g., see
 \cite{roth_coding}).  There are a few cases when optimal codes in terms of the Griesmer bound can be found.  The following codes are optimal in the sense of achieving the Griesmer bound with equality.

\vspace{0.3cm}
\noindent\textit{\textbf{1) Simplex Codes}}\\
 First note that if $d_{\min}(\cC) = 2^{(K-1)B}$, then
 \begin{align*}
 \sum_{i=0}^{K-1}  \frac{ 2^{(K-1)B}} { 2^{iB}} &= 1 + 2^{B}+\cdots+2^{(K-1)B}\\
  &= \frac{2^{KB}-1}{2^B-1}\\
  &=N
 \end{align*}
The \textit{simplex code}, which is the dual code of the Hamming code, can achieve this minimum Hamming distance.
 If we denote $GF(2^B)$ as $\{0,1,2,\cdots,q-1\},$ a generator matrix of the $2^B$-ary simplex code is given as
 \begin{equation}\label{G_simplex}
 \bG_{simplex} = \left[
\begin{array}{ccccccc}
0 &0  &\cdots &0   & \cdots& 1& 1   \\
 0 &0& \cdots   & 0 & & q-1&q-1   \\
\vdots  &\vdots  &  & \vdots& \reflectbox{$\ddots$}  &\vdots&\vdots\\
 0 &1& \cdots   & 1 & & q-1&q-1\\
    1 & 0 & \cdots & q-1 &\cdots&q-2& q-1
\end{array}
\right].
 \end{equation}
In words, the generator matrix $\bG_{simplex}$ is the $K\times (2^{KB}-1)/(2^B-1)$ matrix with columns chosen to correspond to all non-zero vectors in $GF(2^B)^K$ with first non-zero entry fixed to one.

\vspace{0.3cm}
\noindent \textit{\textbf{2) First-Order Reed-Muller Codes}}\\
A first-order Reed-Muller code exists for $N = 2^{(K-1)B}$ with minimum distance $d_{\min}(\cC) = 2^{(K-2)B}(2^B-1).$    It achieves the Griesmer bound with equality because
\begin{align*}
 &\sum_{i=0}^{K-1}  \left\lceil{\frac{2^{(K-2)B} (2^B-1)}{2^{iB}}}\right\rceil\\
  &\quad= \left\lceil\frac{2^B-1}{2^B}\right\rceil  +(2^B-1) +\cdots+2^{(K-2)B}(2^B-1)\\
  &\quad= 1 + {2^{(K-1)B}-1}\\
  &\quad=N.
\end{align*}
This code has a generator matrix of
\begin{equation}\notag
  \bG_{RM1} = \left[
\begin{array}{cccccccc}
1 & 1 & \cdots & 1 &1 & \cdots &1 & 1\\
0 &0  &\cdots &0 &0   & \cdots& q-1& q-1   \\
 0 &0& \cdots   &0 & 0 & & q-1&q-1   \\
\vdots  &\vdots  &\cdots  &\vdots& \vdots& \reflectbox{$\ddots$}  &\vdots&\vdots\\
 0 &0& \cdots   &0 & 1 & & q-1&q-1\\
    0 & 1 & \cdots & q-1 &0 &\cdots&q-2& q-1
\end{array}
\right].
 \end{equation}
 This corresponds to all possible vectors in $GF(2^B)^{K-1}$ with a one appended to the top of the vector.

\vspace{0.3cm}
 \noindent\textit{\textbf{3) Shortened Concatenated Repetition-Simplex (SCRS) Codes}}\\
 A simple approach to code design when $N\neq 2^{(K-1)B}$ and $N\neq (2^{KB}-1)/(2^B-1)$ is to shorten a concatenated code consisting of a shorter  simplex code and a  repetition code.   In this case, the outer code is the simplex code and the inner code is a repetition code.

To construct our code, we first define two variables
\begin{equation*}
   N_{out}=\frac{(2^{KB}-1)}{(2^B-1)},\quad N_{in} =  \left\lceil \frac{N(2^B-1)}{(2^{KB}-1)}\right\rceil,
\end{equation*}
and construct the $K\times N_{out}N_{in},$ generator matrix
 \begin{align*}
 \bG_{concat} &=  \mathbf{1}_{1\times N_{in}} \otimes \bG_{simplex}\\
 & = \left[
\bG_{simplex} ~
\bG_{simplex} ~
\cdots~\bG_{simplex}
\right]
 \end{align*}
where $\bG_{simplex}$ is the $K\times N_{out}$ simplex code's generator matrix given in \eqref{G_simplex}, $\mathbf{1}_{1\times N_{in}}$ is the $N_{in}$ row vector of all ones (i.e., $\mathbf{1}_{1\times N_{in}} = [1 ~1\cdots 1]$), and $\otimes$ represents the Kronecker product.
If we let
\begin{equation*}
N^{'}=N_{out}N_{in} - N,
\end{equation*}
then the extended code uses the shortened generator matrix given by
 \begin{equation}\notag
 \bG_{extend} =  \bG_{concat}\left[
 \begin{array}{c}
 \bI_{ N}\\
 \b0_{N^{'}\times N}
 \end{array}\right]
 \end{equation}
where $\bI_N$ is the $N\times N$ identity matrix and $\b0_{N^{'}\times N}$ is the $N^{'}\times N$ all zero matrix.

\subsection{SCRS Codes Analyses}
A SCRS code achieves a minimum distance of
\begin{equation}\label{scrs_dmin}
d_{\min}(\cC) \geq  \left\lfloor \frac{N(2^B-1)}{(2^{KB}-1)}\right\rfloor 2^{(K-1)B}.
\end{equation}
When $N = K (2^{KB}-1)/(2^B-1)$, the code is optimal with respect to the Griesmer bound because
\begin{align*}
\sum_{i=0}^{K-1} \left\lceil \frac{N(2^B-1)2^{(K-1)B}}{2^{iB}(2^{KB}-1)}\right\rceil &= \sum_{i=0}^{K-1} K  \frac{2^{(K-1)B}}{2^{iB}} \\
& = K \sum_{i=0}^{K-1}   {2^{iB}} \\
&=  K \left(\frac{2^{KB}-1}{2^B-1}\right)\\
&=N.
\end{align*}
For arbitrary $N$, the following lemma states that the SCRS codes are optimal in terms of the Griesmer bound when $K=2$.
\begin{lemma}
The length $N$ SCRS code with $K=2$ formed from concatenating the $2^B$-ary simplex code and repetition code
has the following properties:\\
\noindent 1) The minimum Hamming distance becomes
\begin{equation*}
d_{\min}(\cC) = \alpha 2^B + r-1
\end{equation*}
where $\alpha = \left\lfloor N/(2^{B}+1)\right\rfloor$, $N = \alpha (2^{B}+1) + r$, and $r$ is the remainder when $N$ is divided by $N_{out} =(2^{B}+1)$.  \\
\noindent 2) The code achieves the Griesmer bound with equality.
\end{lemma}
\begin{IEEEproof}
For any length $N = \alpha N_{out} + r$, the generator matrix can be written as
\begin{equation*}
\bG  = [\underbrace{
\bG_{simplex} ~
\bG_{simplex} ~
\cdots~\bG_{simplex}}_{\alpha} ~\bG_{K\times r}
]
\end{equation*}
where $\bG_{simplex}$ is $K \times N_{out}$ matrix given as
\begin{equation}\notag
\bG_{simplex} = \left[
\begin{array}{cccccc}
1 & 0 & 1 & 1 &\cdots & 1\\
0 & 1 & 1 & 2 & \cdots & q-1
\end{array}\right],
\end{equation}
where the matrix $\bG_{K\times r}$ consists of the first $r$ columns of $\bG_{simplex}.$
The minimum distance of this code is
\begin{equation}\notag
d_{\min}(\cC) =   \alpha 2^{B} + d_{K\times r}
\end{equation}
where $d_{K\times r}$ is the minimum distance of the code with generator matrix $\bG_{K\times r}.$

It is obvious that $d_{K\times r} = 0$ if $r=0,1$ and $d_{K\times r} = 1$ if $r=2.$  For more general $r,$ the $K\times r$ code has a $(r-K)\times r$ parity check matrix
\begin{equation}\notag
\left[
\begin{array}{cccccc}
1 & 0 & \cdots & 0 & 1 & 1\\
0 & 1 & \ddots & 0 & 1 & 2\\
\vdots & & \cdots & \vdots & \vdots & \vdots\\
0& 0& \cdots & 0 & 1 & r-3  \\
0& 0& \cdots & 1 & 1 & r -2
\end{array}\right].
\end{equation}
By checking the minimum number of columns in the parity check matrix for which a nontrivial combination gives the all zero out, we can see that
$d_{K\times r} = r-1$ for $r>0.$
Therefore, $d_{\min}(\cC) = \alpha 2^B + r-1.$

Note that the Griesmer bound for $K=2$ tells us that to provide a  minimum distance of $d$ requires  a code of length at least $d +  1$ when $d=1,2,\ldots,2^B.$  This means that our $K\times r$ code is optimal in the sense of achieving equality in the Griesmer bound.    For the entire code, note that
\begin{align*}
\alpha 2^B + r-1 + \left\lceil\frac{ \alpha 2^B + r-1}{2^B} \right\rceil &= \alpha 2^B + r-1  + \alpha + 1 \\
&= \alpha (2^B +1) + r\\
& = N.
\end{align*}
This shows that the SCRS codes are optimal in terms of the Griesmer bound.
\end{IEEEproof}

Note that the case when $K=2$ is a very practical scenario in distributed reception.  For example, the scenario corresponds to the case when the transmitter sends 16QAM (or QPSK) symbol and each receive node forwards QPSK (or BPSK) symbol to the fusion center.  It would be very unlikely for the receive nodes (that might consists of cheap sensors) to send a high-order modulation symbol than QPSK to the fusion center in distributed reception.

\textbf{Remark:} The proposed SCRS codes are suitable to selected subset ML decoding explained in Section \ref{decoding_schemes} because every $N_{out}$-th receive node shares a common processing rule in the SCRS codes.

\subsection{Achievable Rate}\label{achievable_rate}
The achievable rate (or the mutual information) given channel realizations contained in the vector $\bh$ is given by
\begin{equation*}
I(\bh) = \sum_{s\in\cS}\sum_{\bu\in\cU} Pr(s)Pr(\bu\mid s,\bh) \log_2\left(\frac{Pr(\bu\mid s,\bh)}{Pr(\bu\mid\bh)}\right)
\end{equation*}
where $\cU$ denotes the set of all $2^{NB}$ possible outputs from the receive nodes.  In most communication systems, the source can be modeled well as uniformly distributed over $\mathcal{S}$, which simplifies the mutual information to
\begin{align*}
I_{u}(\bh) &= \frac{1}{M}\sum_{s\in\cS}\sum_{\bu\in\cU}\left\{ Pr(\bu\mid s,\bh)\vphantom{\log_2\left(\frac{Pr(\bu\mid s,\bh)}{\frac{1}{M}\sum_{s'\in\cS}Pr(\bu\mid s',\bh)}\right)}\right.\\
 &\qquad \qquad \times \left.\log_2\left(\frac{Pr(\bu\mid s,\bh)}{\frac{1}{M}\sum_{s'\in\cS}Pr(\bu\mid s',\bh)}\right)\right\}.
\end{align*}
Given this, the average achievable rate is given by
\begin{equation}\label{Ravg}
R_{avg} = E\left[I_u(\bh)\right]
\end{equation}
with the expectation taken with respect to $\bh$.

Note that all of these achievable rate expressions are dependent on the quantization structure used at each receive node.  This is implicit because the transition probabilities between the input symbols and output symbols are dependent on this quantization structure.  However, in general, it is hard to derive transition probabilities analytically, which prevents to have a closed-form expression of the achievable rate of the proposed coded diversity technique. Thus, we numerically study the achievable rate of the proposed coded diversity technique in Section \ref{sec6} and show that the proposed scheme can provide benefits even with respect to the achievable rate in some scenarios.

\section{Numerical Studies}\label{sec6}
We perform Monte-Carlo simulations to evaluate the proposed coded diversity technique in this section.  We assume all channel entries are independent, Rayleigh distributed, i.e., $h_i\sim \cC\cN(0,1)$ for all $i$, during simulation; however, the proposed techniques can be applied to any kind of channel models of interest.  The proposed scheme is based on the SCRS codes to simulate different numbers of the receive node $N$.

We first compare the proposed coded diversity technique to the scheme from \cite{code_dist_detec1}.  In \cite{code_dist_detec1}, the optimized codeword set matrix for local decision and decoding rules using simulated annealing for QPSK constellation data symbols, $B=1$ processing at each receive node, and $N=10$ nodes is given as\footnote{The concept of the codeword set matrix is similar to a code for the constellation points $\cS$ in \eqref{proposed_code_matrix}.  Both matrices represent the constellation points $\cS$.  However, local decision rules are completely different, i.e., the local decision rules in \cite{code_dist_detec1} are based on the codeword set matrix while the proposed scheme relies on the generator matrix $\bG$.}
\begin{equation*}
  \text{Codeword Set Matrix:}~(6,12,4,9,12,9,12,6,1,3).
\end{equation*}
Each integer in the matrix represents binary column vector of the matrix, e.g., the integer 12 in column 2 represents $[0~0~1~1]^T$.  Each row and column of the matrix represents one of QPSK constellation point and decision rule of each receive node, respectively.  For example, if node 2 (which corresponds to column 2 in the codeword set matrix) detects the transmitted symbol as the first or second (third or fourth) QPSK constellation points, it forwards 0 (1) to the fusion center.  With $N$ binary bits forwarded from all the receive nodes, the fusion center adopts the same decoding rule with the proposed coded diversity technique for the fair comparison.
\begin{figure}[t]
  \centering
  \includegraphics[width=0.9\columnwidth]{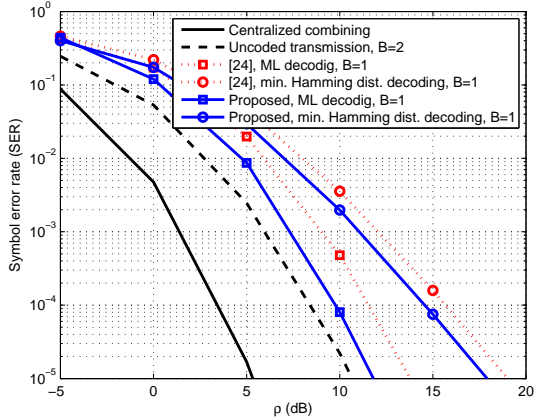}\\
  \caption{Symbol error rate (SER) vs. SNR in dB scale with $M=4$ and $N=10$.  Each receive node of the proposed scheme and the scheme from \cite{code_dist_detec1} forwards $B=1$ bit per channel use to the fusion center while uncoded transmission relies on $B=\log_2 M$ forwarded bits per channel use from each node.}\label{M4B1}
\end{figure}
\begin{figure}[t]
  \centering
  \includegraphics[width=0.9\columnwidth]{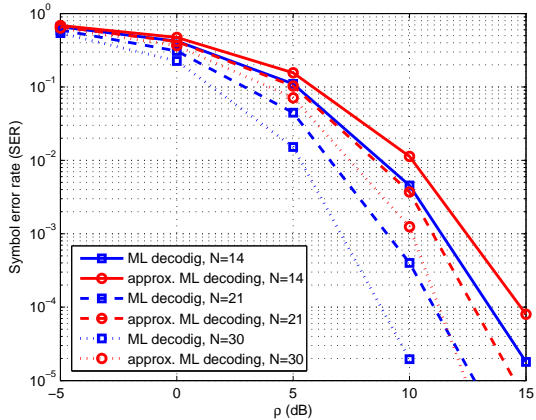}\\
  \caption{Symbol error rate (SER) of the proposed coded diversity technique with ML and selected subset ML decoding schemes according to SNR in dB scale.  $M=8$ and $B=1$.}\label{ML_vs_apporx_ML}
\end{figure}

Fig. \ref{M4B1} compares the symbol error rate (SER) of the proposed scheme (using the SCRS codes) and the scheme in \cite{code_dist_detec1} for QPSK constellation data symbols according to the transmit SNR $\rho$ with $N=10$ receive nodes.  We also plot the results of centralized combining with $\bz=\bh/\|\bh\|$ in \eqref{cas} and uncoded $B=\log_2 M$ bits transmission from each receive node to the fusion center for comparison purpose.  In uncoded transmission, the fusion center perform majority decoding based on the forwarded $N$ estimated symbols from the receive nodes.

In both ML decoding and minimum Hamming distance decoding, the proposed scheme outperforms the scheme in \cite{code_dist_detec1}.\footnote{We do not consider selected subset ML in this case because selected subset ML is not suitable to the scheme in \cite{code_dist_detec1}.}  It is expected to have these results because the corresponding SCRS code has the minimum Hamming distance of 6 while the scheme in \cite{code_dist_detec1} has the minimum Hamming distance of 5.  According to Lemma \ref{lemma_soft} and \ref{lemma_hard}, the diversity orders of the SCRS code are 6 and 3 for ML and minimum Hamming distance decoding, respectively, while those of the scheme in \cite{code_dist_detec1} are 5 and 3, respectively.  The results in Fig. \ref{M4B1} perfectly match with the derivations of Lemma \ref{lemma_soft} and \ref{lemma_hard}.  Note that the SCRS code with ML decoding shows the same diversity order with uncoded transmission with much less transmission overhead from the receive nodes to the fusion center.  Because we have an explicit expression of the SCRS code for an arbitrary number of the receive nodes $N$, the proposed coded diversity technique is very practical and easy to implement.

We compare the proposed diversity technique with ML and selected subset ML decoding schemes in Fig. \ref{ML_vs_apporx_ML}.  We set $M=8$ and $B=1$ (which gives $N_{out}=7$ for the SCRS code) with different numbers of the receive nodes $N$.  When $N_{out}$ (or $L$ with the notation in the selected subset ML decoding section) divides $N$, it is clear that selected subset ML has the same diversity order with ML decoding although selected subset ML suffers from a certain SNR loss.  Note that even when $N_{out}$ does not divide $N$ (the case when $N=30$ in Fig. \ref{ML_vs_apporx_ML}), selected subset ML gives comparable diversity gain with ML decoding with much less complexity.

\begin{figure*}
\centering
\subfloat[$M=8$ (8PSK) constellation for $\cS$ and $B=1$.]{
\includegraphics[width=0.9\columnwidth]{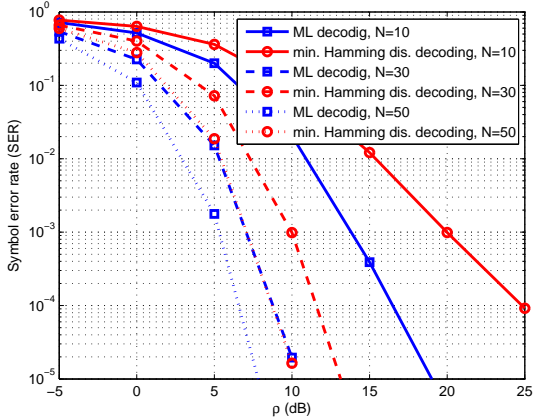}
\label{M8B1_fig}
}
\subfloat[$M=16$ (16QAM) constellation for $\cS$ and $B=2$.]{
\includegraphics[width=0.9\columnwidth]{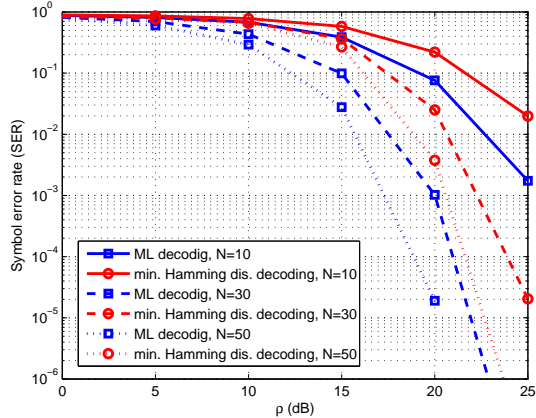}
\label{M16B2_fig}
}\\
\caption{Symbol error rate (SER) vs. SNR in dB scale with different values of $M$, $B$, and $N$.}
\label{ser_fig}
\end{figure*}

In Figs. \ref{M8B1_fig} and \ref{M16B2_fig}, we plot SER of the proposed coded diversity technique according to $\rho$ with different values of $M$, $B$, and $N$.  We can see from the figures that as the number of the receive nodes increases, we have better SER with the same $\rho$.  Moreover, the number of the receive nodes $N$ does not need to be large to achieve practical SER of $10^{-2}$ or $10^{-3}$ with moderate $\rho$ for all cases, which clearly shows the practicality of the proposed coded diversity technique.

Finally, we perform simulations to verify the average achievable rate of the proposed scheme which is explain in Section \ref{achievable_rate}.  We compare $R_{avg}$ in \eqref{Ravg} of the proposed coded diversity technique, centralized combining, and uncoded transmission.  To simplify simulations, we set $\cS$ with QPSK constellation, $B=1$, and $N=3$, which is the same setup as the motivating example in Section \ref{sec2}.  We consider two different scenarios, i.e., 1) Rayleigh fading channels for all channels between the transmitter and the receive nodes, 2) normalized Rayleigh fading channels such that channel amplitudes are normalized as $|h_1|=|h_3|=1.5$ and $|h_2|=0.3$ for all channel realizations.  The second scenario would be the case when the second node is in a deep fade while two other nodes are in stably good channel conditions.

\begin{figure*}
\centering
\subfloat[Rayleigh fading for all channels.]{
\includegraphics[width=0.9\columnwidth]{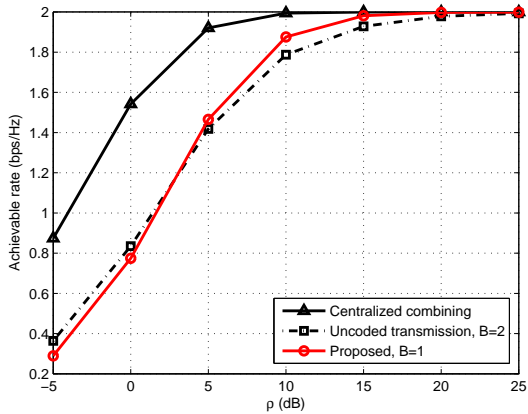}
\label{rate_rayleigh}
}
\subfloat[Fading channel with normalized channel gain of $|h_1|=|h_3|=1.5$ and $|h_2|=0.3$.]{
\includegraphics[width=0.9\columnwidth]{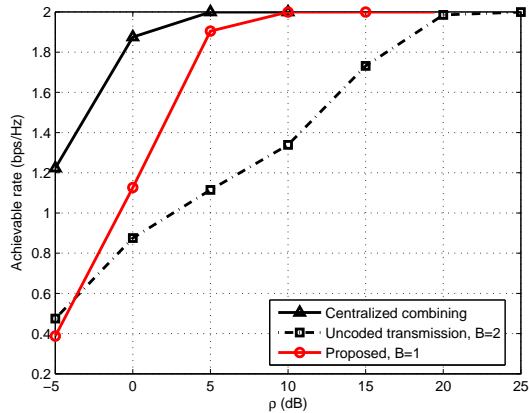}
\label{rate_fixed}
}\\
\caption{Achievable rate vs. SNR in dB scale with $M=4$, $N=3$, and $B=1$.}
\label{ach_rate_fig}
\end{figure*}

We plot the results of the scenarios 1 and 2 in Figs. \ref{rate_rayleigh} and \ref{rate_fixed}, respectively.  In the first scenario, the proposed scheme and uncoded transmission are comparable with each other.  This results are reasonable because the proposed coding structure is not intended to increase the achievable rate.  However, the proposed scheme outperforms uncoded transmission in the second scenario.  This is because the second node that processes the imaginary component of the transmitted symbol is in a deep fade in uncoded transmission, resulting in significant achievable rate degradation.  On the contrary, the proposed coded diversity is even better in the second scenario than the first since the fusion center can obtain much of mutual information only from node 1 and 3 that are in good channel conditions.

\section{Conclusion}
We proposed a unified framework for coded diversity distributed reception in this paper.  We consider distributed reception for the case when a transmitter broadcasts a signal to multiple geographically separated receive nodes through fading channels, and each receive node processes and forwards the received signal to a fusion center.  The fusion center then tries to detect the transmitted signal exploiting the forwarded data from all the receive nodes and channel state information if available.  The proposed coded diversity technique is based on the strong connection between the distributed reception problem and coding problem in coding theory.  By leveraging this connection, we are able to adopt appropriate linear block codes, e.g., simplex and first-order Reed-Muller codes that achieve the Griesmer bound with equality, to design processing rules at the receive nodes and maximize the diversity gain.  We also developed novel shortened concatenated repetition-simplex (SCRS) codes to support an arbitrary number of the receive nodes.  We analytically proved that the SCRS codes are optimal with respect to the Griesmer bound in many practical scenarios.  We also evaluated the proposed coded diversity technique by numerical studies.  Because of its simple and flexible structure, the proposed technique can be applied to various scenarios including cellular systems, wireless sensor networks, and radar systems.

The proposed coded diversity technique only can support an integer value of compression ratio.  Supporting an arbitrary value of compression ratio is a nontrivial problem, and it would be an interesting future research topic to generalize the proposed framework in this direction.

\bibliographystyle{IEEEtran}
\bibliography{refs_coded_div}

\end{document}